\documentclass[twocolumn,prb,showpacs,preprintnumbers,amsmath,amssymb]{revtex4}

\usepackage[dvips]{graphicx}
\usepackage{dcolumn}
\usepackage{color}
\usepackage{bm}
\newcommand{\CoCo}{Ca$_3$Co$_2$O$_6$}
\newcommand{\CoRh}{Ca$_3$CoRhO$_6$}
\newcommand{\FeRh}{Ca$_3$FeRhO$_6$}

\begin{document}
\title{Insulating state and the importance of the spin-orbit coupling in Ca$_3$CoRhO$_6$}

\author{Hua Wu, Z. Hu, D. I. Khomskii, and L. H. Tjeng}
  \address{ II. Physikalisches Institut, Universit{\"a}t zu K{\"o}ln,
   Z{\"u}lpicher Str. 77, D-50937 K{\"o}ln, Germany}


\begin{abstract}
We have carried out a comparative theoretical study of the
electronic structure of the novel one-dimensional {\CoRh} and
{\FeRh} systems. The insulating antiferromagnetic state for the
{\FeRh} can be well explained by band structure calculations with
the closed shell high-spin $d^5$ (Fe$^{3+}$) and low-spin
$t_{2g}^{6}$ (Rh$^{3+}$) configurations. We found for the {\CoRh}
that the Co has a strong tendency to be $d^7$ (Co$^{2+}$) rather
than $d^6$ (Co$^{3+}$), and that there is an orbital degeneracy
in the local Co electronic structure. We argue that it is the
spin-orbit coupling which will lift this degeneracy thereby
enabling local spin density approximation + Hubbard U (LSDA+U)
band structure calculations to generate the band gap. We predict
that the orbital contribution to the magnetic moment in {\CoRh} is
substantial, i.e. significantly larger than 1 $\mu_B$ per formula
unit. Moreover, we propose a model for the contrasting intra-chain
magnetism in both materials.
\end{abstract}


\maketitle

\section{Introduction}

The synthesis of the one-dimensional cobaltate Ca$_3$Co$_2$O$_6$
\cite{Fjellvag96,Aasland97} and the discovery of the exotic
stair-step jumps in the magnetization at regular intervals of the
applied magnetic
field \cite{Kageyama97a,Maignan00} have triggered a flurry of
research activities, both experimentally
\cite{Maignan00,Kageyama97b,Martinez01,Raquet02,Ray03,Hardy03,
Maignan03,Maignan04,Samp04,Flahaut04,Sekimoto04,Hardy04,Takubo05}
and theoretically.
\cite{Whangbo03,Vidya03,Eyert04,Fresard04,Dai05,Villesuzanne05,
Wu05,Kudasov06}
This compound consists of [Co$_2$O$_6$]$_\infty$ chains running
along the $c$ axis of the hexagonal unit cell, with in each chain
alternating CoO$_6$ octahedra and CoO$_6$ trigonal 
prisms.\cite{Fjellvag96} The intra-chain coupling is ferromagnetic
(FM) with a Curie temperature of $T_C = 24\rm~K$ and the
inter-chain antiferromagnetic (AF) with a N\'{e}el temperature of
$T_N = 10\rm~K$.\cite{Kageyama97a} Owing to an 
inter-chain magnetic frustration associated with the triangular 
lattice, the magnetic ground-state is either a partially disordered
AF state or a spin-freezing state.\cite{Kageyama97a} 
Substitution of the Co by
other transition metals results in quite dramatic changes of the
properties. The {\CoRh} compound has its $T_C$ shifts up to a
high value of 90 K and $T_N$ to 35 K,
\cite{Niitaka99,Niitaka01a,Niitaka01b,Samp02,Davis03,Loewenhaupt03}
but the type of its magnetic ground-state is similar to 
Ca$_3$Co$_2$O$_6$,\cite{Niitaka01a,Niitaka01b} while {\FeRh} is an intra-chain AF with 
$T_N = 12\rm~K$.\cite{Niitaka99,Davis03,Niitaka03} 
In these compounds the Rh
ions occupy the octahedral sites while the Co/Fe reside within
the trigonal prisms.\cite{Niitaka99,Davis03}

There is a debate about the electronic structure of these
one-dimensional cobaltates in the literature, in particular for
{\CoCo}
\cite{Fjellvag96,Aasland97,Kageyama97a,Kageyama97b,Maignan00,Martinez01,Raquet02,
Ray03,Hardy03,Maignan03,Maignan04,Samp04,Flahaut04,Sekimoto04,Hardy04,Takubo05,
Whangbo03,Vidya03,Eyert04,Fresard04,Dai05,Villesuzanne05,Wu05}
and {\CoRh}.\cite{Takubo05,Whangbo03,
Niitaka99,Niitaka01a,Niitaka01b,Samp02,Davis03,Loewenhaupt03,Eyert05}
Consensus has yet to arrive about the valence, spin and orbital
state of the Co ions, without which one could not make a reliable
modeling of the magnetic properties. Focusing on the {\CoRh}
system, it was originally thought that the valence state is
Co$^{2+}$/Rh$^{4+}$ based on magnetization measurements and bond
valence sum.\cite{Niitaka99,Niitaka01a} Subsequent neutron
studies concluded, however, that the Co$^{3+}$/Rh$^{3+}$ scenario
is a better description.\cite{Niitaka01b,Loewenhaupt03} Very
recently, a photoemission investigation favors yet the original
Co$^{2+}$/Rh$^{4+}$ assignment.\cite{Takubo05} Existing band
structure calculations also provide a mixed message: a
generalized-gradient-approximation (GGA) study suggests the
Co$^{3+}$/Rh$^{3+}$ state,\cite{Whangbo03} while an LSDA
calculation leaves this issue open.\cite{Eyert05} Important is to
note that both calculations predict {\CoRh} to be a metal, in
strong disagreement with the 
experiment.\cite{Maignan03,Takubo05}

To resolve these disagreements, we carried out a theoretical study
of the electronic structure and magnetic properties of {\CoRh} in
which we took the insulating state of the material
\cite{Maignan03,Takubo05} as a key finding. For this we applied the LSDA+U
method \cite{Anisimov91,Anisimov93} in order to take into account more
explicitly the correlated motion of the electrons typical for
transition metal oxides. Using {\FeRh} as a reference, we find
that the Co ions have a lower valence than Fe, and that the Co
are in an orbitally degenerate state while the Fe are not. The
valence assignment coming from our calculations is the
Fe$^{3+}$/Rh$^{3+}$ for {\FeRh} and Co$^{2+}$/Rh$^{4+}$ for
{\CoRh}. The insulating state for {\FeRh} can be readily
reproduced, but for {\CoRh} we infer that the spin-orbit coupling
(SOC) must be included in the LSDA+U scheme. We make a testable
prediction, namely that there is a large orbital contribution to
the magnetic moment in {\CoRh}, about 1.7 $\mu_B$ per formula
unit. We also calculate the various exchange constants and
propose a model to explain the weak intra-chain AF in {\FeRh} and 
enhanced intra-chain FM in {\CoRh}.

\begin{figure}[h]
 \centering\includegraphics[width=8cm]{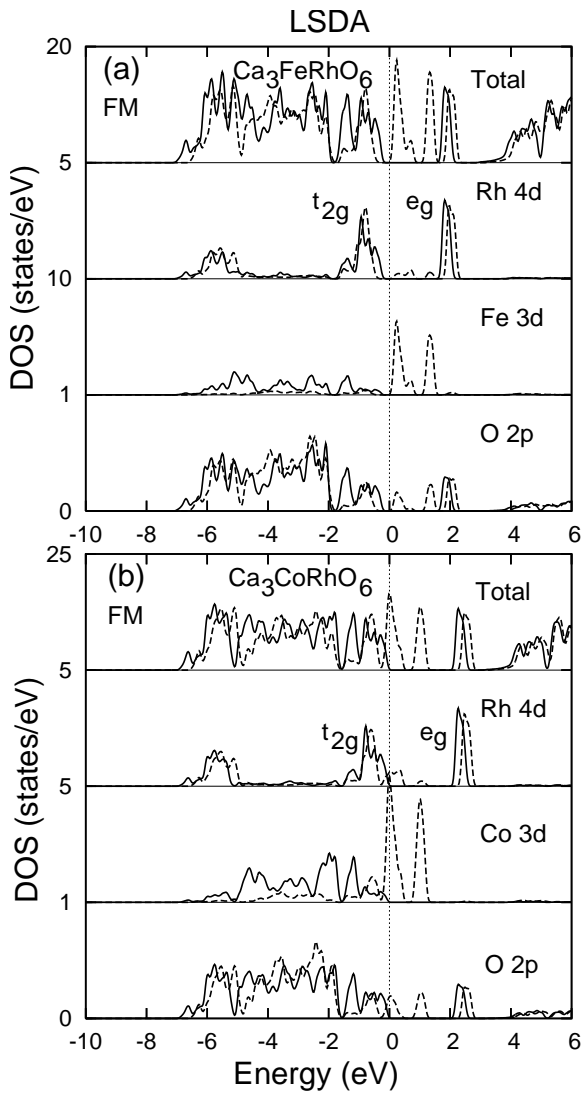}
 \caption{Density of states (DOS) of {\FeRh} (panel a) and {\CoRh}
 (panel b) in the ferromagnetic (FM) state from LSDA. The Fermi
 level is set at zero energy. Both panels show, from top to bottom,
 the total DOS, the Rh $4d$, Fe/Co $3d$, and O $2p$ partial DOS.
 Solid (dashed) lines depict the spin-up (down) states.}
 \label{fig1}
\end{figure}

\section{Results and Discussion}

Our calculations are performed by using the full-potential
augmented plane waves plus local orbital method.\cite{WIEN2k} We
took the crystal structure data determined by x-ray diffraction.
\cite{Niitaka99} The muffin-tin sphere radii are chosen
to be 2.7, 2.0 and 1.6 Bohr for Ca, Co/Fe/Rh, and O atoms,
respectively. The cut-off energy of 16 Ryd is used for plane wave
expansion of interstitial wave functions, and 5$\times$5$\times$5
{\bf k}-mesh for integrations over the Brillouin
zone, both of which ensuring a sufficient numerical accuracy. The
SOC is included by second-variational method with the scalar
relativistic wave functions,\cite{WIEN2k} and actually 
$l_z$ and $s_z$ are 
good quantum numbers due to the special trigonal crystal field
of the Co sites as seen below. The easy magnetization
direction is along the $c$-axis chains. 
In view of those $c$-axis chains being well separated
in the hexagonal $ab$-plane and of these two compounds behaving
like a quasi-one-dimensional system, we study in this work the type and
origin of the intra-chain magnetic coupling, thereby assuming
a FM inter-chain coupling.

As a reference, we first calculate the electronic structure of
{\FeRh} in the FM state using the LSDA. Fig. 1(a) shows the
density of states (DOS). One can clearly see that the octahedral
Rh $4d$ has a large $t_{2g}$-$e_g$ crystal-field (CF) splitting of
about 3 eV.\cite{note1} 
With the $t_{2g}$ shell completely full and the $e_g$
completely empty, the Rh is formally 3+ ($4d^6$) and takes the
low-spin (LS, $S$=0) state. One can also observe that the Fe $3d$
spin-up states are essentially completely full and the spin-down
completely empty. The Fe is thus formally 3+ ($3d^5$) and
high-spin (HS, $S$=5/2). The Fe partial DOS also shows that the
unoccupied spin-down states are split roughly into two groups
with about 1 eV separation. This is caused by the presence of a
trigonal CF, making ($d_1$, $d_{-1}$) orbitals to lie higher than
the nearly degenerate ($d_0$, $d_2$, $d_{-2}$).
\cite{Aasland97,Whangbo03,Vidya03,Eyert04,Fresard04,
Dai05,Villesuzanne05,Wu05}
Covalency reduces the Fe moment to 3.73 $\mu_B$ but creates at
the same time 0.20 $\mu_B$ moment at the Rh and 0.13 $\mu_B$ at
each O, so that the total moment is still 5.00 $\mu_B$, see Table
I. The material is an insulator, since it is effectively a closed
shell system due to the sufficiently large CF splitting at the Rh
and exchange splitting on the Fe. We have also calculated the
{\FeRh} in the intra-chain AF state, and found very similar
DOS'ses (the spin part not considered). Important is that the
total energy of the AF is lower than the FM by about 11 meV (see
Table I), explaining why {\FeRh} is an AF. Our LSDA finding 
is in close agreement with the GGA results of Villesuzanne and
Whangbo.\cite{Villesuzanne05}

Fig. 1(b) depicts the LSDA results for the {\CoRh} in the FM
state. The DOS shows quite a number of similarities with that of
the {\FeRh}, including the large CF at the Rh and the
characteristic trigonal CF splitting at the Fe/Co. The only
significant difference is the fact that the Co $3d$ lies lower in
energy than the Fe $3d$, which is reasonable since Co is less
electropositive than Fe. The consequences are, however, quite
dramatic: the Fermi level now straddles through the ($d_0$, $d_2$,
$d_{-2}$) part of the Co $3d$ spin down band, meaning that
{\CoRh} would be a metal. This LSDA result is in clear disagreement
with experimental observations.\cite{Maignan03,Takubo05} Also Whangbo
\textit{et al.} \cite{Whangbo03} found a metallic solution. We
have investigated the intra-chain AF state for {\CoRh} and found
that it is also metallic. It has, however, a higher energy than
the FM by about 106 meV, see Table I. This means that system
tends to be FM, at least if one is willing to trust these LSDA
results as far as the magnetism is concerned. It is obvious,
nevertheless, that this half-metallic solution is an artifact of
the LSDA.

\begin{figure}[h]
 \centering\includegraphics[width=8cm]{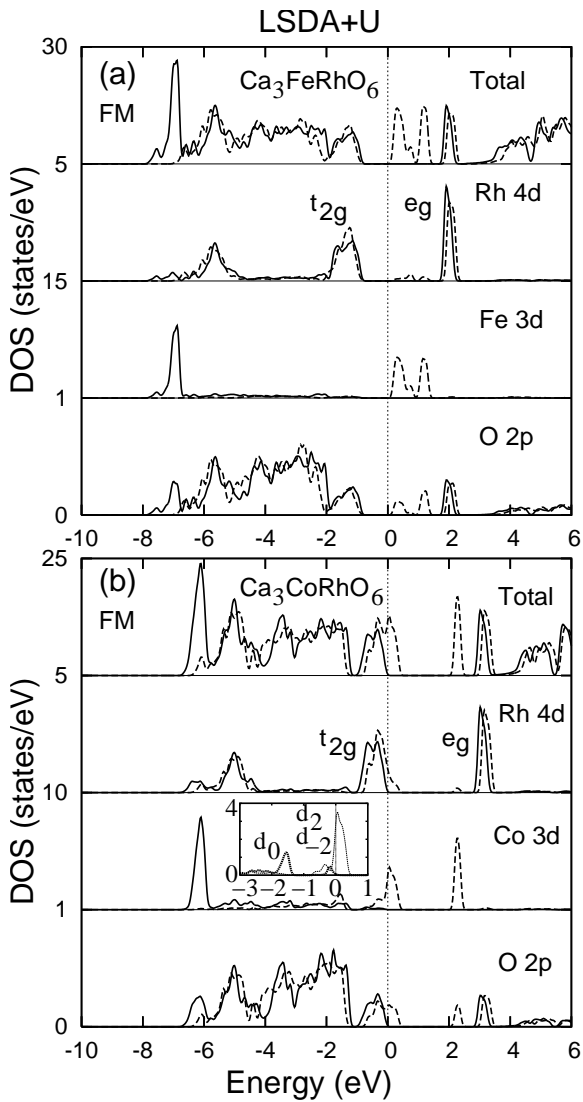}
 \caption{Density of states (DOS) of {\FeRh} (panel a) and {\CoRh}
 (panel b) in the ferromagnetic (FM) state from LSDA+U.
 Inset: close up of the down-spin Co $d_0$, $d_2$ and $d_{-2}$ DOS.
 See Fig. 1 caption for other notations.}
 \label{fig2}
\end{figure}

The LSDA of {\CoRh} shows that the Fermi level is located in a
narrow Co $3d$ band with no more than 1 eV width. It is
then also natural to expect that already modest electron
correlation effects at the Co sites will have a pronounced effect
on the energetics of this band. We therefore carried out LSDA+U
band structure calculations \cite{Anisimov93} for both the {\CoRh}
and {\FeRh} with the Coulomb energy $U$ is set to 5, 4, and 3 eV, and
the Hund's exchange parameter $J_H$ to 0.9, 0.9, 0.5 eV for the
Co $3d$, Fe $3d$, and Rh $4d$ shells, respectively. 

We will first discuss the results for the {\FeRh}, our reference system.
Fig. 2(a) shows the LSDA+U DOS of {\FeRh} in the FM state. One can
observe that the energy separation between the fully occupied
$t_{2g}$ shell and the fully unoccupied $e_g$ has increased in
comparison with the LSDA results [Fig. 1(a)], in accordance with
the inclusion of $U$. This also happens with the separation
between Fe $3d$ spin-up and spin-down states, whereby the
spin-up states are pushed down to about --7 eV.
This deep-lying state reflects
the high stability typical for $3d^5$ ions, in which the high-spin configuration
of the half-filled shell gives an energy gain of 4$J_H$. With the Fe spin-up
states being at the bottom of the valence band, the 
covalency with the Rh $4d$ and O $2p$ bands is much reduced. 
This is then also   
reflected in the increase (with respect to LSDA) of the Fe moment
to 3.95 $\mu_B$ and decrease of the Rh to 0.11 $\mu_B$ and of each O
to 0.11 $\mu_B$. The total magnetic moment should not change: it
remains indeed at 5.00 $\mu_B$, see table I. As far as the low
energy scale physics is concerned, the band gap has increased
from about 0.2 eV to an appreciable 0.9 eV. It does not
directly scale with the $U$'s, since it is determined by the
occupied Rh $t_{2g}$ and the unoccupied Fe $3d$ spin-down states.
Note that our conclusion that {\FeRh} is an insulator 
with HS-Fe$^{3+}$/LS-Rh$^{3+}$ and is a Heisenberg
spin-chain system (see below), is $U$-independent.

Calculations for the intra-chain AF state of {\FeRh} give very similar 
DOS'ses (the spin part not considered), with practically the same total energy 
as the FM (the difference being about 2 meV is within the error bar,
see Table I). 
This and the above LSDA result indicate that the
intra-chain AF exchange interaction is very weak.
In this situation the effects not included in our calculations, such as
the detailed type of inter-chain ordering (we assumed that spins in all 
neighboring chains are in phase, i.e., we considered FM inter-chain ordering)
may start to play a role and may finally determine the type of long-range
ordering in {\FeRh}. Thus from the present LSDA+U calculations we can only 
conclude that the FM and AF intra-chain orderings in this system are practically 
degenerate.
We present below (Section III) qualitative arguments that in reality most 
probably the intra-chain interaction in {\FeRh} is weak but AF. All this 
is not inconsistent with the rather small $T_N$ value of 12 K.

The LSDA+U results for {\CoRh} are shown in Fig. 2(b). Similar to
the {\FeRh} case, the inclusion of $U$ increases the splitting
between the Rh $4d$ $t_{2g}$ and $e_g$ orbitals as well as Co $3d$
spin-up and spin-down bands. Also similar is the reduction of the
covalency of the $3d$ spin-up as it is shifted to very low
energies. Surprisingly, however, the insulating state is
\textit{not} formed. The Fermi level still straddles through the
lower-energy part of the Co $3d$ spin-down band. The
main influence of $U$ here is only that the effective crystal
splitting with the ($d_1$, $d_{-1}$) orbitals is increased. To
investigate why there is no gap opening, we plot in the inset of
Fig. 2(b) a close up of the down-spin Co $d_0$, $d_2$ and $d_{-2}$ 
states in
the vicinity of the Fermi level. We can now observe that the
spin-down $d_0$ is fully occupied, and that it is the $d_2$ and
$d_{-2}$ bands which are partially occupied. This means first of
all, with the spin-up states fully occupied, that the Co ion is
essentially in the $d^7$ or 2+ valence state, and not in the
$d^6$ or 3+ as proposed recently.
\cite{Whangbo03,Niitaka01b,Loewenhaupt03} Secondly, the inset
reveals that it is the degeneracy of the $d_2$ and $d_{-2}$ bands
which makes the LSDA+U to be inoperative to open the gap. If the
$d_0$ were higher in energy than the $d_2$ and $d_{-2}$, or if
the Co valence were 3+, then LSDA+U would certainly have produced
a band gap. For completeness, we have also calculated the AF
state with the LSDA+U, and found also a metallic state with a 13
meV higher total energy, see Table I.

We propose to include the SOC to lift the $d_2$,$d_{-2}$
degeneracy. The use of the SOC to lift the degeneracy in
correlated insulators has surprisingly been done only in a few
instances.\cite{Solovyev98,Kunes03} In most cases, this has been
omitted, even for materials for which it is known that the
orbital contribution to the magnetic moment is substantial.
\cite{Anisimov91,Mazin97,Leonev04,Jeng04,Cococcioni05} We claim
that the SOC issue is essential for this particular {\CoRh}
compound, motivated also by the report that the magnetocrystalline
anisotropy is significant.\cite{Niitaka01a} 
Note that for the Co$^{2+}$ ions in {\CoRh},
the $d_0$ and $d_2$/$d_{-2}$ states are 
split off from the higher-lying $d_1$/$d_{-1}$ states by 
about 1 eV [Fig. 1(b)], i.e., much larger than the SOC energy scale
of about 70 meV.\cite{footnote2} Therefore, the SOC Hamiltonian can be simplified into
just $\zeta$$l_zs_z$ by neglecting the $l_{+}s_{-}$ and
$l_{-}s_{+}$ mixing terms, both of which cause a mixing between
the $d_0$ and $d_1$/$d_{-1}$, or between the $d_2$ ($d_{-2}$) and
$d_1$ ($d_{-1}$). Thus, the $l_z$ and $s_z$ are good quantum numbers
in this particular case.

The results of the
LSDA+U+SOC calculations for {\CoRh} in the intra-chain FM state are 
shown in Fig. 3. We now can observe that a gap has been opened near the
Fermi level, consistent with the experimental finding that the
system is an insulator.\cite{Maignan03,Takubo05} 
The band gap is small
and is given mainly by the weakly split Rh $4d$ bands. The system
is a Mott-Hubbard insulator in which the U in the Rh $4d$ shell
is  the determining factor.\cite{ZSA} As far as the Co $3d$ states are
concerned, the band gap is about 2 eV, i.e. so large that it
no longer determines the band gap of the system. A clearer view
is offered in the inset of Fig. 3 which depicts a close up of the
down-spin Co $d_0$, $d_2$ and $d_{-2}$ states in the vicinity of the 
Fermi level. One can see that the spin-down $d_2$ band is now fully
occupied and the $d_{-2}$ fully unoccupied, i.e. there is no
partial occupation anymore like in the LSDA+U. Interesting is to
note that the spin-down $d_0$ band has also an unoccupied
component slightly above the Fermi level, but not at the Fermi
level. This can be attributed to the band formation with the
unoccupied Rh $4d$ $3z^{2}-r^{2}$ state (i.e. $a_{1g}$ in the
trigonal CF), demonstrating the Rh$^{4+}$ valence state with the
$t_{2g}^{5}$ configuration and the one-dimensional character along
the $c$-axis chain. 

\begin{table*}[ht]
\caption{Calculated electronic states of 
{\FeRh} and {\CoRh} (FM:
ferromagnetic; AF: antiferromagnetic; -I:insulator; -M:metal);
total energy (meV) per formula unit relative to the ground state;
spin moments ($\mu_B$) at each Fe/Co, Rh, and O ion, as well as
in the interstitial region and total magnetic moment per formula
unit. Only the Co has the orbital moment and is shown. The Ca
spin moment is about 0.01 $\mu_B$ and the Fe orbital moment 0.02
$\mu_B$, and are thus omitted.} \label{TableI}

\begin{tabular} {l@{\hskip0.4cm}c@{\hskip0.4cm}c
@{\hskip0.4cm}c@{\hskip0.4cm}c
@{\hskip0.4cm}c@{\hskip0.4cm}c@{\hskip0.4cm}c@{\hskip0.4cm}c
@{\hskip0.4cm}c} \\ \hline\hline

Ca$_3$FeRhO$_6$&state&energy&Fe& & Rh & O & interstitial & total
& Figure\\ \hline
LSDA&FM-I&11&3.73&&0.20&0.13&0.27&5.00&Fig. 1(a) \\
LSDA&AF-I&0&$\pm$3.72&&0&$\pm$0.12&0&0& \\ \hline
LSDA+U&FM-I&0&3.95&& 0.11&0.11&0.23 &5.00&Fig. 2(a) \\
LSDA+U&AF-I&2&$\pm$3.95&& 0&$\pm$0.11&0&0& \\
\hline\hline

Ca$_3$CoRhO$_6$&state&energy&Co &Co$^{orb}$ & Rh & O & interstitial &
total & Figure\\ \hline
LSDA&FM-M&0&2.64&&0.37&0.13&0.16&4.00&Fig. 1(b) \\
LSDA&AF-M&106&$\pm$2.63&&0&$\pm$0.12&0&0& \\ \hline
LSDA+U&FM-M&0&2.85&&0.23&0.13&0.15&4.00&Fig. 2(b) \\
LSDA+U&AF-M&13&$\pm$2.87&&0&$\pm$0.11&0&0& \\ \hline
LSDA+U+SOC&FM-I&0&2.72&1.69& 0.54&0.09&0.18 &5.69&Fig. 3 \\
LSDA+U+SOC&AF-I&70&2.73, -2.53&1.69, -1.70& 0.55&0.09, -0.01&0.07&1.00& \\
\hline\hline
\end{tabular}
\end{table*}

\begin{figure}[ht!]
 \centering\includegraphics[width=8cm]{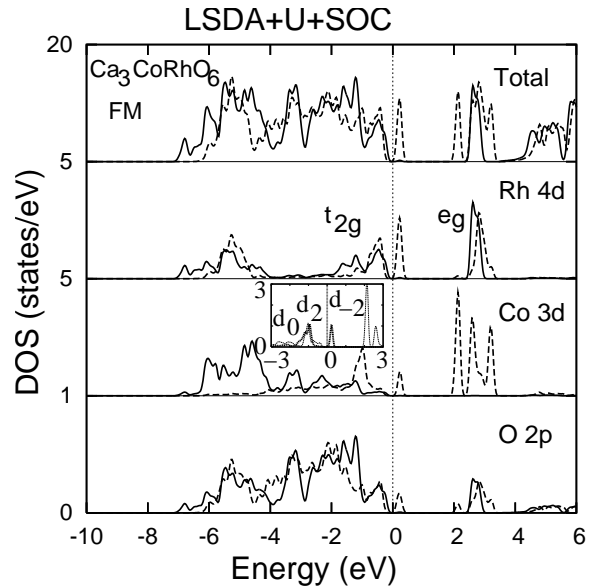}
 \caption{Density of states (DOS) of {\CoRh}
 in the ferromagnetic (FM) state from LSDA+U+SOC.
 Inset: close up of the down-spin Co $d_0$, $d_2$ and $d_{-2}$ DOS.
 See Fig. 1 caption for other notations.}
 \label{fig3}
\end{figure}

While the approach to include the SOC is sound in view of the
sizable magnetocrystalline anisotropy,\cite{Niitaka01a} a
definitive justification needs to come from a determination of the
magnitude of the orbital contribution to the magnetic moment: our
LSDA+U+SOC calculations ($U$=5 eV for Co) predict for the 
Co$^{2+}$ ions not only a spin moment of 2.72
$\mu_B$ but especially a very large orbital moment of 1.69
$\mu_B$ along the $c$-axis chain. This is a testable prediction,
which is actually confirmed by a recent
x-ray magnetic circular dichroism study.\cite{Hu06}
We verified that both the calculated spin and orbital moments 
stay constant
within 0.1 $\mu_B$ when the $U$ for Co is varied in the range of 2.5-6 eV. 
This huge orbital moment and the SOC firmly fix the 
orientation of
the total magnetic moment along the $c$-axis chain, and thus the 
magnetism of {\CoRh} is highly Ising-like. The
calculations also predict an appreciable 0.54 $\mu_B$ spin moment
on the Rh$^{4+}$ $4d$, and 0.09 $\mu_B$ on each oxygen, showing the
effect of covalency.
Note that the Rh$^{4+}$ ions carry no orbital moment,
because the one hole resides on the $a_{1g}$ singlet split off from 
the $t_{2g}$ manifold by the trigonal CF.
We also note that
the Co$^{2+}$/Rh$^{4+}$ valence state of Ca$_3$CoRhO$_6$, as we 
obtained in this work, remain unchanged by counting the electron 
occupation of the down-spin Co $d_0$ and $d_2$ orbitals (the up-spin Co $d$
shell is fully occupied) and the hole occupation of the down-spin 
Rh $d_0$ (i.e. $a_{1g}$) orbital, once an insulating gap is opened at 
$U>$2.5 eV for Co and $U>$2 eV for Rh 
in our LSDA+U calculations including the spin-orbit coupling. 
All in all, our Co$^{2+}$/Rh$^{4+}$ solution
is thus consistent with the large total magnetic moment as
measured by neutron diffraction.\cite{Niitaka01a,Niitaka01b,Loewenhaupt03}

We also have carried out LSDA+U+SOC calculations for {\CoRh} in
the intra-chain AF state, and we have found that this state is 70
meV higher in energy than the FM state, see Table I. This allows
us to estimate the effective FM exchange parameter of a
Co$^{2+}$ pair being 15 meV, using the simple Heisenberg model
with the HS-Co$^{2+}$ spin-3/2. These numbers are substantially
larger than for the {\CoCo} system: there the AF-FM difference is
12 meV and the exchange parameter about 1.5 meV.\cite{Wu05} Our
calculations thus nicely explain why the intra-chain $T_C$ of
{\CoRh} (90 K) is much higher than that of {\CoCo} (24 K). For
completeness, we have also carried out LSDA+U+SOC calculations
for the {\FeRh}. The results, however, are almost identical to
those of the LSDA+U, since the SOC is not operative for the closed
spin-up Fe$^{3+}$ $3d$ shell and the full Rh$^{3+}$ $t_{2g}^{6}$
state. The absence of an orbital moment accounts for its
Heisenberg spin character and very weak magnetocrystalline
anisotropy.\cite{Niitaka03}

\section{Model of Magnetism}

In order to explain both the experimentally observed and
theoretically (computationally) confirmed weak intrachain AF
($T_{\rm N}$=12 K) in Ca$_3$FeRhO$_6$ and relatively strong FM
($T_{\rm C}$=90 K) in Ca$_3$CoRhO$_6$, we propose the following
model. Starting from our finding that Ca$_3$FeRhO$_6$ has the HS
Fe$^{3+}$ and LS Rh$^{3+}$ (non-magnetic $t_{2g}^{6}$) valence
state, a charge fluctuation will create the Fe$^{2+}$/Rh$^{4+}$
pair rather than Fe$^{4+}$/Rh$^{2+}$, since it is easier to
stabilize a Rh$^{4+}$ rather than an Fe$^{4+}$. In Fig. 4(a), we
sketch the normal superexchange mechanism
Fe$^{3+}$--(O$^{2-}$)--Rh$^{3+}$--(O$^{2-}$)--Fe$^{3+}$ via the
intermediate non-magnetic Rh$^{3+}$ (in the figure the O$^{2-}$ is
omitted for clarity). Since the HS Fe$^{3+}$ has a closed up-spin
$3d$ shell, only the down-spin Rh$^{3+}$ electron [e.g., the
$a_{1g}$ (3z$^2$--r$^2$)] can make a virtual excitation, which
naturally explains the intra-chain AF, with a low $T_{\rm N}$ due
to the large Fe-Fe distance of 5.39 \AA~. In fact, this result is
a direct analog of the first Goodenough-Kanamori-Anderson 
rule.\cite{GKA}

\begin{figure}[h]
 \centering\includegraphics[width=9cm]{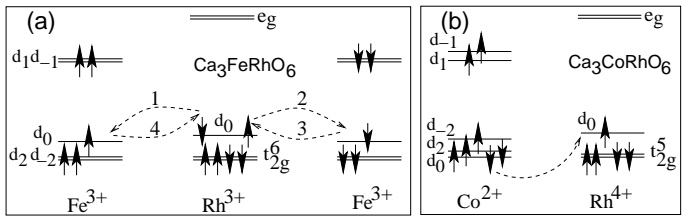}
 \caption{Model of the intra-chain AF in Ca$_3$FeRhO$_6$ (a)
and of the intra-chain FM in Ca$_3$CoRhO$_6$ (b).}
 \label{fig4}
\end{figure}

Ca$_3$CoRhO$_6$ in contrast, has the HS Co$^{2+}$ and LS Rh$^{4+}$
($t_{2g}^{5}$, $S$=1/2) valence state. We therefore consider the
Co$^{3+}$/Rh$^{3+}$ charge fluctuation (rather than the
Co$^{1+}$/Rh$^{5+}$ with a rarely observed Co$^{1+}$ state) for
the virtual excitation path Co$^{2+}$--(O$^{2-}$)--Rh$^{4+}$.
Since Rh$^{4+}$ has a large $t_{2g}$-$e_g$ CF splitting which
makes the empty $e_g$ states to be costly to reach, it will be the
$t_{2g}^{5}$ states which will allow for an electron transfer to
the empty down-spin $a_{1g}$ orbital, as shown in Fig. 4(b). If
the Co$^{2+}$ and Rh$^{4+}$ have a FM alignment, the down-spin
Co$^{2+}$ $a_{1g}$ ($d_0$) electron will do so; otherwise (in the
AF Co$^{2+}$/Rh$^{4+}$ alignment) the spin-up $a_{1g}$ electron
will do, but obviously with a higher energy cost than the former
by 3 times Hund's exchange energy of the Co ion. Thus the
Co$^{2+}$ and Rh$^{4+}$ strongly prefer the FM alignment, being
an analog of the second Goodenough-Kanamori-Anderson 
rule.\cite{GKA} 
In addition to the spin-triplet
oxygen-hole mediated intra-chain FM ordering present in both
Ca$_3$CoRhO$_6$ and Ca$_3$Co$_2$O$_6$,\cite{Wu05} this mechanism accounts
for a higher $T_{\rm C}$ in the former than in the latter. Note
that if Ca$_3$CoRhO$_6$ had the HS Co$^{3+}$ and LS non-magnetic
Rh$^{3+}$ valence states, as in the case of Ca$_3$Co$_2$O$_6$
with the HS trigonal Co$^{3+}$ and LS non-magnetic octahedral
Co$^{3+}$, its $T_{\rm C}$ would be lower than that of
Ca$_3$Co$_2$O$_6$ due to a longer distance between the magnetic
HS Co$^{3+}$ ions in the former than latter.

\section{Conclusion}

In summary, we have performed systematic LSDA and LSDA+U
calculations with the inclusion of the spin-orbit coupling for the
new quasi-one-dimensional spin-chain materials {\FeRh} and {\CoRh}. 
We conclude that {\CoRh} is a Mott insulator with the
high-spin Co$^{2+}$ and low-spin Rh$^{4+}$ configurations, and
that the correlated insulator {\FeRh} has the closed-shell
high-spin Fe$^{3+}$ and low-spin Rh$^{3+}$. We predict that
{\CoRh} has a very large orbital moment at the Co$^{2+}$ site,
which also explains the strong magnetocrystalline anisotropy and
the highly Ising-spin like behavior. The inclusion of 
the spin-orbit coupling in the present LSDA+U calculations is also
crucial to obtain the insulating state. This highlights
the importance of the spin-orbit
coupling in (nearly) degenerate correlated systems.
 In contrast, {\FeRh} has
spin-only moments and behaves like a Heisenberg spin-chain system.
Moreover, our calculations reproduce the relatively strong
ferromagnetic intra-chain coupling in {\CoRh} but the weak, 
presumably antiferromagnetic one in {\FeRh}, and we propose a model 
to explain the contrasting magnetism.

We thank M. W. Haverkort and T. Burnus for valuable discussions.
This work is supported by the Deutsche Forschungsgemeinschaft
through SFB 608 and by the European project COMEPHS.


\begin{thebibliography}{99}
\bibitem{note1} Actually in both Ca$_3$FeRhO$_6$ and
Ca$_3$CoRhO$_6$ the octahedral Rh feels a trigonal crystal field.
However, the $a_{1g}$ singlet split off from the $t_{2g}$ manifold lies above
the $e_{g}^{\pi}$ doublet only by about 200 (230) meV in Ca$_3$FeRhO$_6$ 
(Ca$_3$CoRhO$_6$), being much less than the cubic $t_{2g}$-$e_g$
splitting of about 3 eV. Therefore, we keep to use in this work
the conventional orbital index $t_{2g}$.
\end{thebibliography}
\end{document}